\documentclass[11pt,twoside]{article}

\usepackage{asp2006}
\usepackage{graphicx}

\begin{document}
\title{The Millennium Galaxy Catalogue: Science highlights}
\author{Simon P. Driver\altaffilmark{1}, Jochen Liske\altaffilmark{2}, and Alister W. Graham\altaffilmark{3}} 
\altaffiltext{1}{SUPA, School of Physics and Astronomy, University of St Andrews, St Andrews, UK}    
\altaffiltext{2}{European Southern Observatory, Karl-Schwarzschild-Str. 2, 85748 Garching, Germany}    
\altaffiltext{3}{Centre for Astrophysics, Swinburne University of Technology, Hawthorn, Australia}    

\begin{abstract} 
The Millennium Galaxy Catalogue (MGC) provides a structural database
comprising 10,095 well resolved galaxies drawn from a 37.5 sq.\ deg
region of sky, with $B_{\mbox{\tiny \sc mgc}} < 20.0$ mag and 96.1 per
cent spectroscopic completeness. The data are being used to
investigate a number of diverse topics including: the nearby galaxy
merger rate (via close pairs and asymmetry); dust attenuation; bulge
and disc luminosity functions; the luminosity size relations; the
supermassive black hole mass function; galaxy bimodality; the
space-density of high and low surface brightness galaxies; blue
spheroids; and the total $B$-band luminosity function. Plans to
extend the MGC in area (200 sq.\ deg), depth ($K_{Vega}=16.5$
mag), resolution (0.5$''$) and wavelength $(u-K)$ are
underway.
\end{abstract}


\section{Introduction}
The Millennium Galaxy Catalogue (MGC; \citealp{mgc1}) originated with
a major award of time with the Wide Field Camera at the Isaac Newton
Telescope. Over a period of three annual cycles (1999-2001) an
equatorial strip 75\ deg long by 0.5\ deg wide was observed via
sequential 12.5 min $B$-band exposures with large overlap regions
suitable for bootstrap calibration. All data were reduced through the
Cambridge Survey Unit Pipeline. Basic photometry was performed using
SExtractor and calibrated as described in \cite{mgc1} providing
photometric and astrometric accuracies of $\pm 0.023$ mag and $\pm
0.08''$ respectively. The catalogue contains over one million galaxies
with 10,095 brighter than $B_{\mbox{\tiny \sc mgc}}=20$ mag over an
effective useable area of 30.88 sq.\ deg, defining {\sc
mgc-bright}. All detections were verified by eye and corrected where
necessary. The MGC overlaps with the 2dFGRS NGP (\citealp{2dfgrs}) and
the SDSS Early Data Release region (\citealp{sdssedr}) providing
redshifts for almost half of {\sc mgc-bright}, the remainder were
obtained over the subsequent three years (2001-2004, see
\citealp{mgc4}) via (predominantly) general user 2dF time at the AAT,
supplemented with additional time at the ANU 2.3m, and the NTT, Gemini
and TNG telescopes to pursue very high and very low surface brightness
systems. Most recently we have completed 2D S\'ersic and S\'ersic
bulge plus exponential disc profiling of all 10,095 galaxies (see
\citealp{mgc8}) with GIM2D (\citealp{gim2d}). All data products were
publicly released at the IAU General Assembly in Prague and are
available via our website: {\bf http://www.eso.org/$\sim$jliske/mgc}. 

\section{Science highlights}

\subsection{The bivariate brightness distribution ($M-<\mu_{eff}>$-plane)}
In \cite{mgc4} we derive the space-density of galaxies as a function
of magnitude {\it and} mean effective surface brightness (the so
called bivariate brightness distribution). Our 2D variant of step-wise
maximum likelihood enables us to compensate for the different volumes
over which galaxies of similar luminosity but distinct surface
brightnesses can be seen, as well as define the MGC selection
boundary. The distribution (see Fig.~1) shows the strong
luminosity-surface brightness relation but more importantly that; (i)
galaxies near $L^*$ ($M_B \sim -19.6$) are well defined by a Gaussian
distribution in surface brightness which is unimpeded by the selection
boundary, and (ii) that faintward of $M_B=-16$ mag the surface brightness
and size selection boundaries start to severely hinder our ability to
sample the dwarf population, causing a loss of $\sim 6$\% of the
dwarfs at the high surface brightness end alone (see
\citealp{mgc6}). The conclusion is that while the bright-end of the
galaxy luminosity function is well defined, the dwarf regime is poorly
constrained and warrants further investigation via a deeper survey
(both in terms of flux limit, spatial resolution and spectroscopic
follow up limit).

\begin{figure}[h]
\vspace{-0.75cm}
\hspace{-1.0cm} \includegraphics[width=9.0cm]{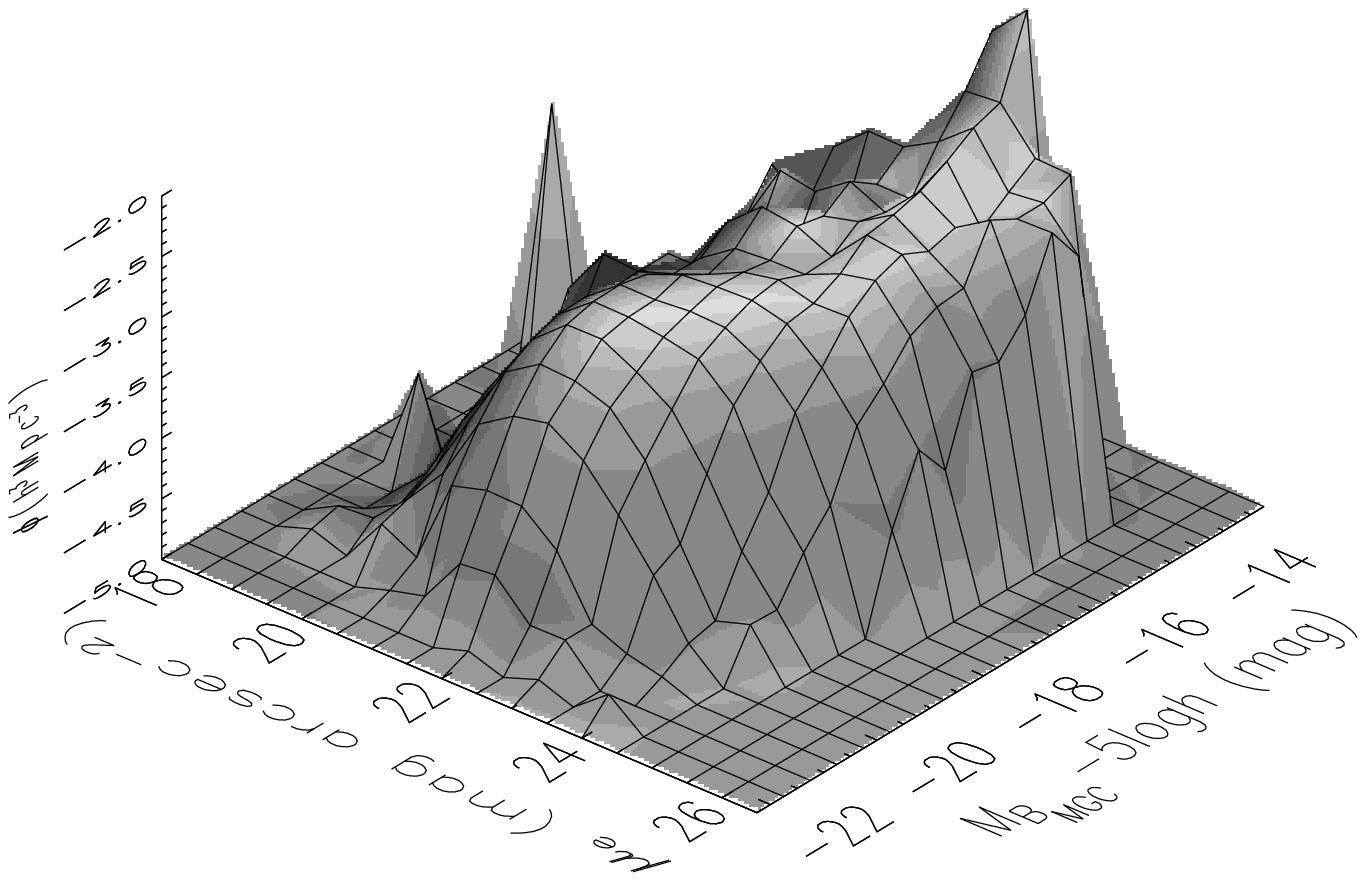}

\vspace{-5.5cm}

\hspace{6.75cm} \includegraphics[width=6.5cm]{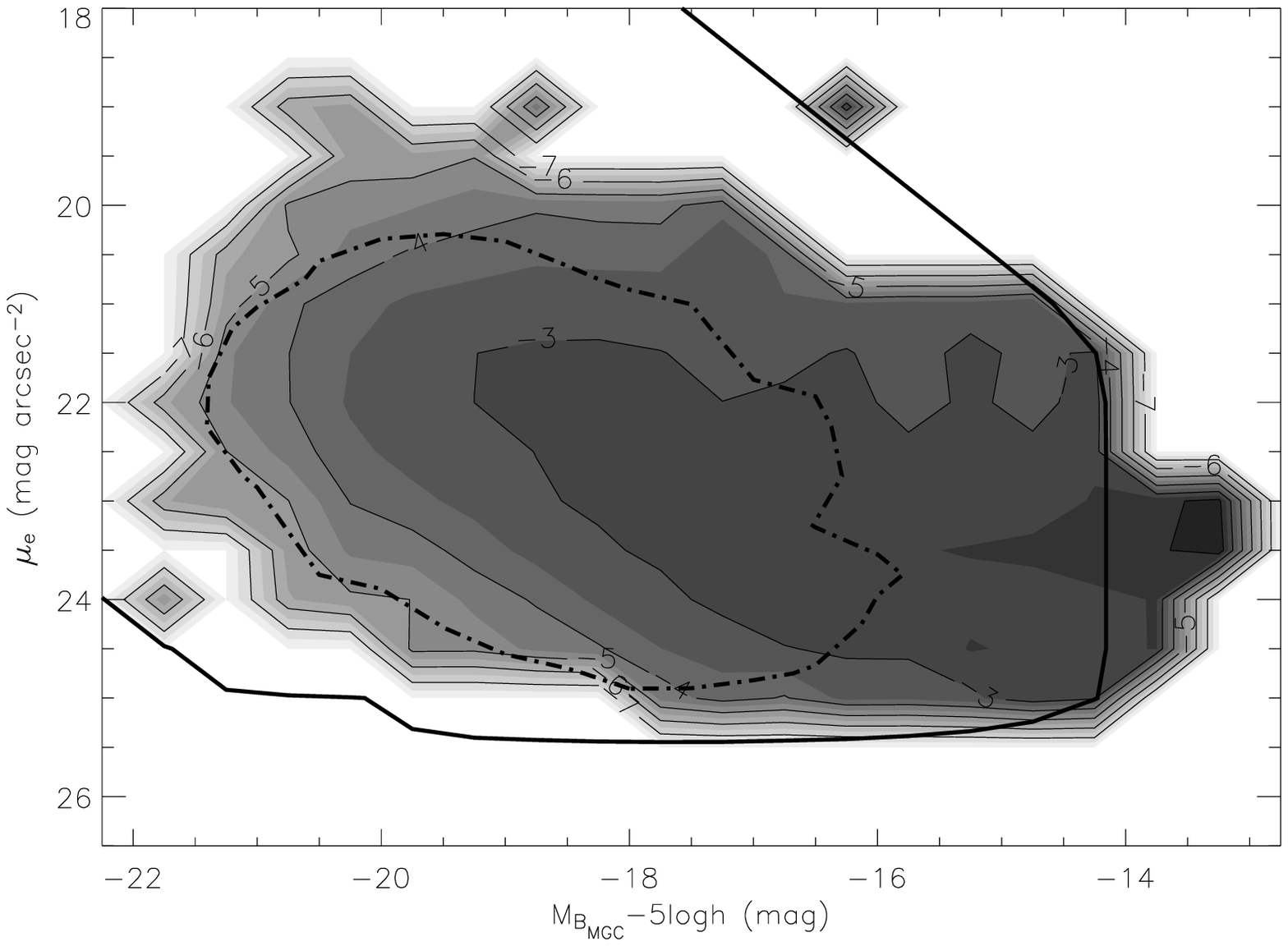}
\caption{The bivariate brightness distribution as a 3D plot ({\it
left}) and a contour plot ({\it right}). The solid line denotes the
MGC selection boundary. Vertical scaling/contours are logarithmic.}
\end{figure}

\subsection{Galaxy bimodality in colour and structure}
In \cite{mgc5} we identified a remarkably strong bimodal distribution
in the colour and structure plane (see Fig.~2). Subdividing by
morphological type we found that ellipticals/lenticulars
(bulge-dominated systems) lie in the red compact peak, late-types
(disc only systems) lie in the blue diffuse peak, and mid-type spirals
(bulge plus disc systems) straddle the two peaks with no obvious
bimodality. We interpreted this to imply that galaxy bimodality is a
fundamental reflection of the two-component nature of galaxies (i.e.,
combinations of spheroids and discs and not two distinct galaxy
types). This in turn implies that galaxies may form via a two stage
process: spheroids form early via a dissipational collapse/rapid merger
phase (suggested by their colours and concentrated 3D structure), and
galaxian discs grow more slowly via splashback (dissipationless
collapse), infall and/or accretion (suggested by their colours and
more fragile 2D structure). In Driver et al. (2007, in prep but see
Driver, Liske \& Graham 2006) we confirm, via our bulge plus disc
structural decompositions of the mid-type spirals, that the
two-component interpretation appears to hold, with the bulges and
discs of mid-type spirals lying in distinct non-overlapping peaks in
the colour-structure plane. The current dominant semi-quiescent phase
is also suggested by our relatively low merger rate found from the MGC
data by both dynamically close pairs (\citealp{rdp1}) and asymmetry
indicators (\citealp{rdp2}).

\begin{figure}[h]
\vspace{-0.5cm}
\includegraphics[width=14.0cm,height=8.0cm]{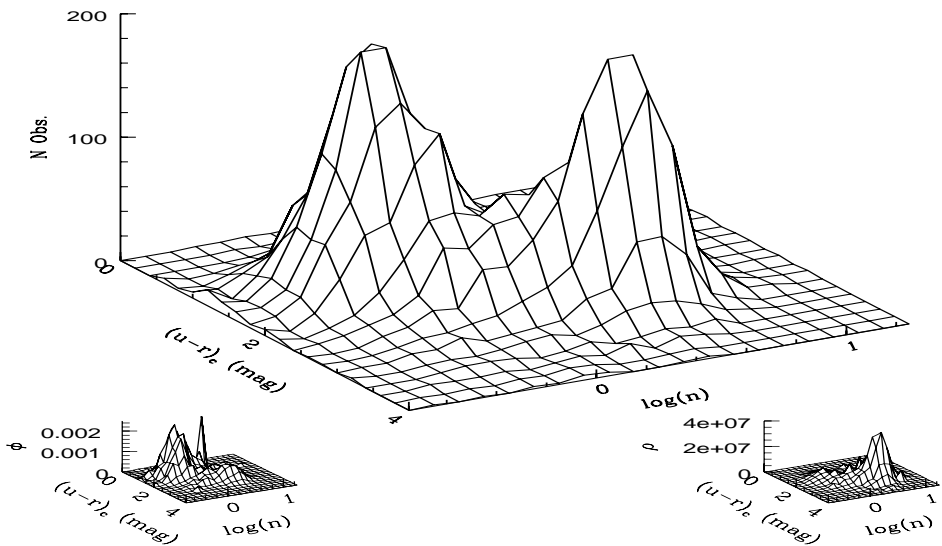}
\caption{The bimodality of galaxies in the colour-structure plane,
defined by observed number, $(u-r)_{\rm core}$ and $\log(n)$ (where
$n$ is the single profile S\'ersic index, see \citealp{sersic}.). The
smaller panels show the same data but with the z-axis changed to
either the volume corrected space-density ({\it left}) or the stellar
mass density ({\it right}).}
\end{figure}

\subsection{The luminosity function of bulges and discs etc.}
In \cite{mgc14} we derive the luminosity function of discs, red
ellipticals, classical bulges, blue ellipticals and pseudo-bulges (see
Fig.~3). We derive their luminosity densities and stellar mass
densities and find the following breakdown in terms of the stellar
mass of 58:13:26:1.5:1.5 (\%) respectively. Adopting the two-mode model for
galaxy formation (rapid bulge assembly followed by semi-quiescent disc
formation) this implies that the later disc-mode is responsible for
the bulk of stars visible today (i.e., $\sim60$\%).

\begin{figure}[h]
\vspace{-7.5cm}
\includegraphics[width=12.0cm,height=14.0cm]{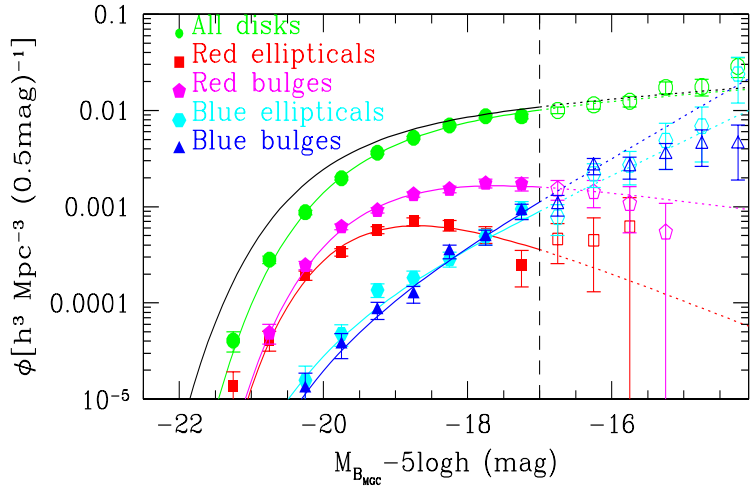}
\caption{The luminosity function of spheroids (red ellipticals+red bulges), discs and blue spheroids (blue ellipticals plus pseudo-bulges).}
\end{figure}

\subsection{Dust attenuation of bulges and discs}
In \cite{mgc15} we derive the bulge and disc luminosity functions in
intervals of inclination and find a strong dependency of the
characteristic turnover luminosity with disc inclination. We interpret
this as due to dust and find our data agree well with the predictions
of \cite{tuffs}.  Using the Tuffs et al.\ models we constrain the mean
face-on opacity of discs and derive a face-on attenuation of 0.2 mag
for galaxy discs and 0.84 mag for galaxy bulges. Together the total
galaxy attenuation is dramatic (see Fig.~4) implying that only 50\% of
the $B$-band photons produced by stars emerge from the nearby galaxy
population with the remainder absorbed by dust grains in the ISM. This
has dramatic implications for the $B$-band luminosity density and
derived $B$-band galaxy luminosity functions, and also affects stellar
masses based on optical colours.

\begin{figure}[h]
\vspace{-7.0cm}
\includegraphics[width=\textwidth,height=13.0cm]{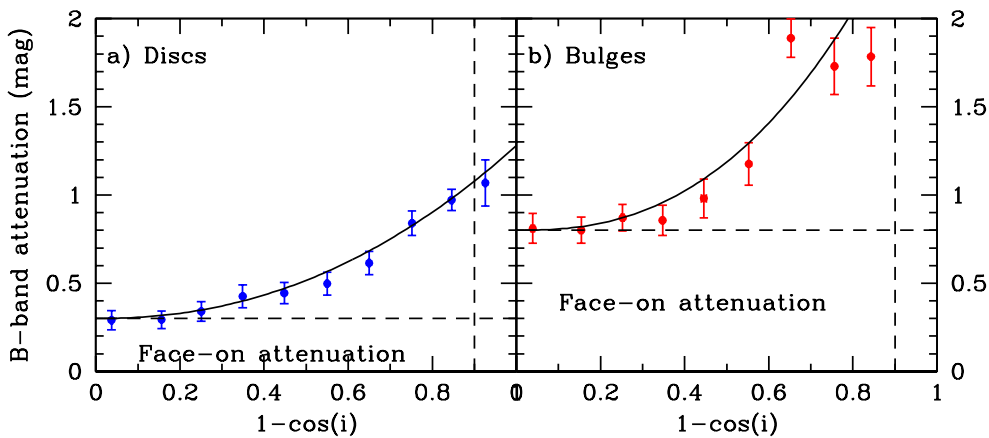}
\caption{The MGC $B_{\mbox{\tiny \sc mgc}}$ band attenuation of discs (left) and bulges (right).}
\end{figure}

\subsection{The $B$-band luminosity function corrected for dust attenuation}
In Driver et al. (2007, in prep.) we rederive the $B$-band galaxy
luminosity function(s) incorporating our dust attenuation correction
(as well as upgrade our photometry from Kron to {\sc gim2d}
magnitudes) and apply individual $k(z)$ and $e(z)$ corrections for the
bulge and disc components (see Fig.~5). We find that the $B$-band galaxy
luminosity function turnover moves brightwards by 1.0
mag when dust is correctly taken into account. Returning to the
component stellar mass contributions we now find that the stellar mass
locked up in bulges is significantly higher and revise our earlier
values to: 58:10:29:1.5:1.5 (see \S 2.3).

\begin{figure}[h]
\vspace{-6.5cm}
\includegraphics[width=14.0cm,height=13.0cm]{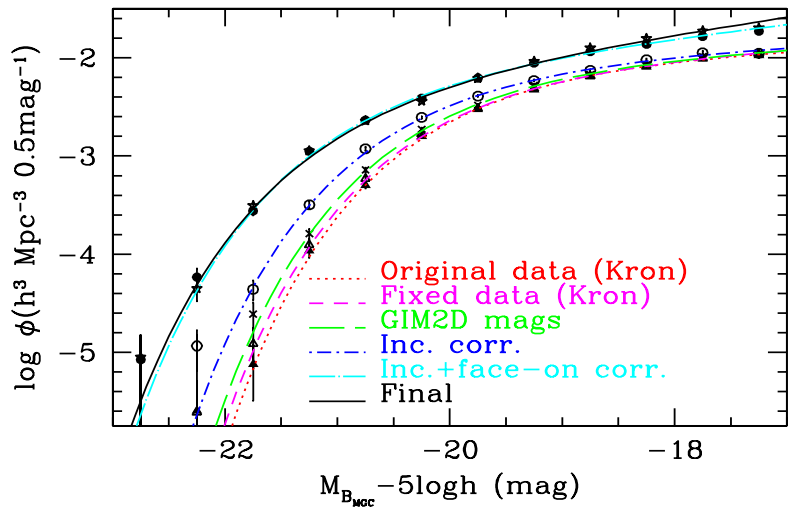}
\caption{The progression of the MGC $B$-band luminosity function.}
\end{figure}

\subsection{Supermassive black holes}
\cite{Mn} provide an updated $M_{\rm bh}$--$n$ relation (akin to the
$M_{\rm bh}$--$\sigma$ relation, see Novak, Faber \& Dekel 2006) which can be
used for predicting accurate supermassive black hole masses in
other galaxies.  Using the measured (spheroid) S\'ersic indices drawn
from the 10k galaxies in the MGC, we provide a new estimate of the local
supermassive black hole mass function (\citealp{mgc9}; see Fig.~6).
The observational simplicity of our approach, and the direct
measurements of the black hole predictor quantity, i.e.\ the S\'ersic
index, for both elliptical galaxies and the bulges of disc galaxies
makes it straightforward to estimate accurate black hole masses in
early- and late-type galaxies alike.  Moreover, the $M_{\rm bh}$--$n$
relation has a total absolute scatter of only 0.31 dex.

\begin{figure}[h]
\vspace{-6.5cm}
\includegraphics[width=\textwidth,height=12.0cm]{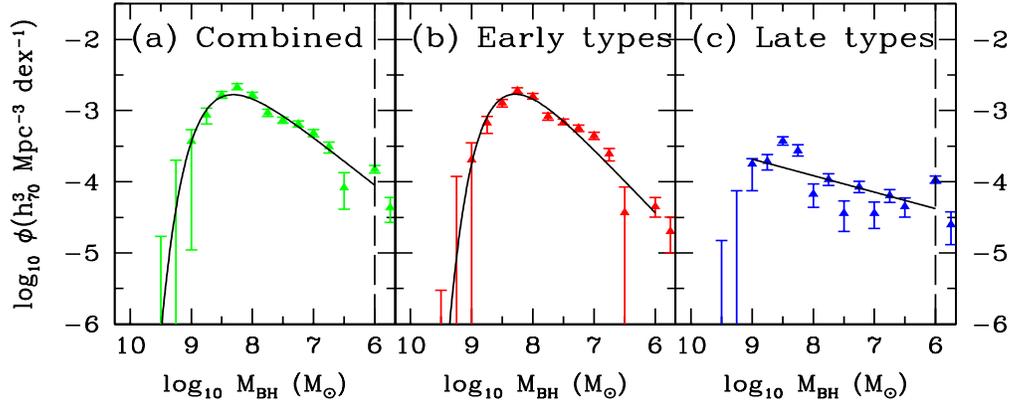}
\caption{The derived MGC supermassive black hole mass functions for all galaxy types ({\it left}), early-types ({\it middle}) and late-types ({\it lower}).}
\end{figure}

\section{Summary}
The MGC is a relatively modest project but is now starting to produce
some fairly unique science. In cosmological terms the survey is
currently too small to place constraints on cosmological parameters
but we can investigate the breakdown of the cosmic baryon budget
today. From the MGC we are finding that $8.3h_{0.7}$\% of the baryons
produced in the Big Bang are currently in stars, $0.006h_{0.7}$ per
cent are in interstellar dust grains, and $0.008h_{0.7}$\% are locked up in
supermassive black holes.  Plans are afoot to now expand the MGC by
ingesting the UKIRT/UKIDSS, VST KIDS and VISTA VIKING data along with
extended spectroscopic coverage with AAOmega at the AAT and supporting
GALEX and Spitzer observations. The extended survey (GAMA), will
provide a $10\times$ gain in sky coverage, $2\times$ gain in spatial
resolution, $2$ mag increase in spectroscopic follow-up, and $4\times$
gain in wavelength coverage and should provide the ultimate post-SDSS
structural and spectroscopic galaxy resource for multipurpose galaxy
studies. One key objective will be to explore CDM on 1 Mpc to 1 kpc
scales where the interplay between CDM and baryon physics becomes
crucial.



\end{document}